\documentclass[nofootinbib,prx,twocolumn,showpacs,superscriptaddress,notitlepage,superscriptaddress,amsmath,longbibliography]{revtex4-2}

\usepackage{lipsum}  
\usepackage{dsfont}
\usepackage{amsmath}
\usepackage{braket}
\usepackage{amssymb}
\usepackage{amsthm}
\usepackage[noend]{algpseudocode}
\usepackage[noend]{algorithm}
\usepackage{siunitx}
\usepackage{amsfonts}
\usepackage{comment}
\usepackage[normalem]{ulem}
\usepackage{graphicx}
\usepackage{color,framed}
\usepackage{hyperref}
\usepackage{enumerate}
\usepackage{lipsum}
\usepackage{slashed}
\usepackage{url}
\usepackage{bbm}
\usepackage{chngcntr}
\counterwithout{equation}{section}
\usepackage{tikz,pgfplots}
\usepackage{pgfplotstable}
\usepackage{siunitx}
\usepackage{comment}
\usepackage{graphicx}

\makeatletter
\newcommand{\setlabel}[1]{\edef\@currentlabel{#1}\label}
\makeatother

\usepgfplotslibrary{fillbetween}

\hypersetup{
    colorlinks=true, 
    linktoc=all,     
    linkcolor=blue,  
}

\def \beq {\begin{equation}}
\def \eeq {\end{equation}}
\def \beqa {\begin{eqnarray}}
\def \eeqa {\end{eqnarray}}
\def \bseq {\begin{subequations}}
\def \eseq {\end{subequations}}

\newcommand{\D}[1]{\text{d}#1}

\pgfplotsset{compat=1.18}
\begin{document}

\title{Spectral functions on a quantum computer through system-environment interaction}

\newcommand{\firstaff}{Quantinuum, Leopoldstrasse 180, 80804 Munich, Germany}
\newcommand{\secondaff}{Quantinuum, Terrington House, 13-15 Hills Road, Cambridge CB2 1NL, UK}
\newcommand{\thirdaff}{Quantinuum, 303 S Technology Ct, Broomfield, CO 80021, USA}

\author{Etienne Granet}
\email{etienne.granet@quantinuum.com}
\affiliation{\firstaff}

\author{Ramil Nigmatullin}
\affiliation{\secondaff}

\author{David T. Stephen}
\affiliation{\thirdaff}

\author{Henrik Dreyer}
\affiliation{\firstaff}

\begin{abstract}
Spectral functions measured with angle-resolved photoemission spectroscopy (ARPES) provide key insight to elucidate the band structure of materials. Comparison with theory requires computing dynamical one-point functions in some equilibrium state, which can be classically challenging. Their measurement on quantum computers poses multiple problems and comes with a large sampling overhead when standard techniques are used. We introduce an efficient way of measuring spectral functions on a quantum computer by directly modeling the interaction of the system with the environment involved in ARPES experiments. We develop quantum circuits whose local expectation values are proportional to the spectral function $A(k,\omega)$ for all momentum $k$ and a specific chosen frequency $\omega$. Although coming with a qubit and two-qubit gate overhead, our approach requires $\mathcal{O}(N)$ times less sampling than previous approaches, translating into a factor $\mathcal{O}(N)$ faster in runtime, and is particularly adapted to ion-trap quantum computers. The algorithm requires to implement a fermionic Fourier transform (FFT). We write out an efficient gate decomposition for generic radix-$n$ FFT and benchmark it on hardware for radix-$3$ on $27$ qubits. We finally demonstrate our algorithm on a Quantinuum System Model H2 ion-trap system, computing the spectral function on a one-dimensional system of $27$ sites, using $54$ qubits.

\end{abstract}

\maketitle

\section{Introduction} One of the main applications of quantum computers is the simulation of many-body quantum physical systems \cite{miessen2023quantum,alexeev2024quantum}, with significant recent progress on multiple quantum computing platforms \cite{kim2023evidence,haghshenas2025digital,abanin2025observation,alam2025programmable,evered2025probing,granet2025superconducting}. In material science, a key experiment to probe the band structure of materials is angle-resolved photoemission spectroscopy (ARPES) \cite{damascelli2003angle,sobota2021angle,zhang2022angle}. Under linear response it is proportional to the spectral function $A(k,\omega)$. This spectral function is defined in terms of the retarded Green function, here written for a one-dimensional system of size $N$
\begin{equation}
    A(k,\omega)=2{\rm Im} \int_{-\infty}^{\infty}\D{t}\sum_{j=1}^N e^{ikj} e^{-i\omega t}G_{\rm ret}(j,t)\,,
\end{equation}
where $G_{\rm ret}(j,t)=i\theta(t)\langle \{c_j^\dagger(t) ,c_0(0)\}\rangle$ with $c_j,c_j^\dagger$ canonical fermionic annihilation and creation operators. This can be rewritten as $A(k,\omega)=A^+(k,\omega)+A^-(k,\omega)$ with
\begin{equation}\label{aa}
\begin{aligned}
    &A^+(k,\omega)=\sum_{j=1}^N\int_{-\infty}^{\infty}\D{t}e^{ikj}e^{-i\omega t} \langle c^\dagger_j(t)c_0(0)\rangle\\
    &A^-(k,\omega)=\sum_{j=1}^N\int_{-\infty}^{\infty}\D{t}e^{-ikj}e^{-i\omega t} \langle c_j(0)c^\dagger_0(t)\rangle\,.
\end{aligned}
\end{equation}

\begin{figure}
\includegraphics[width=0.95\linewidth]{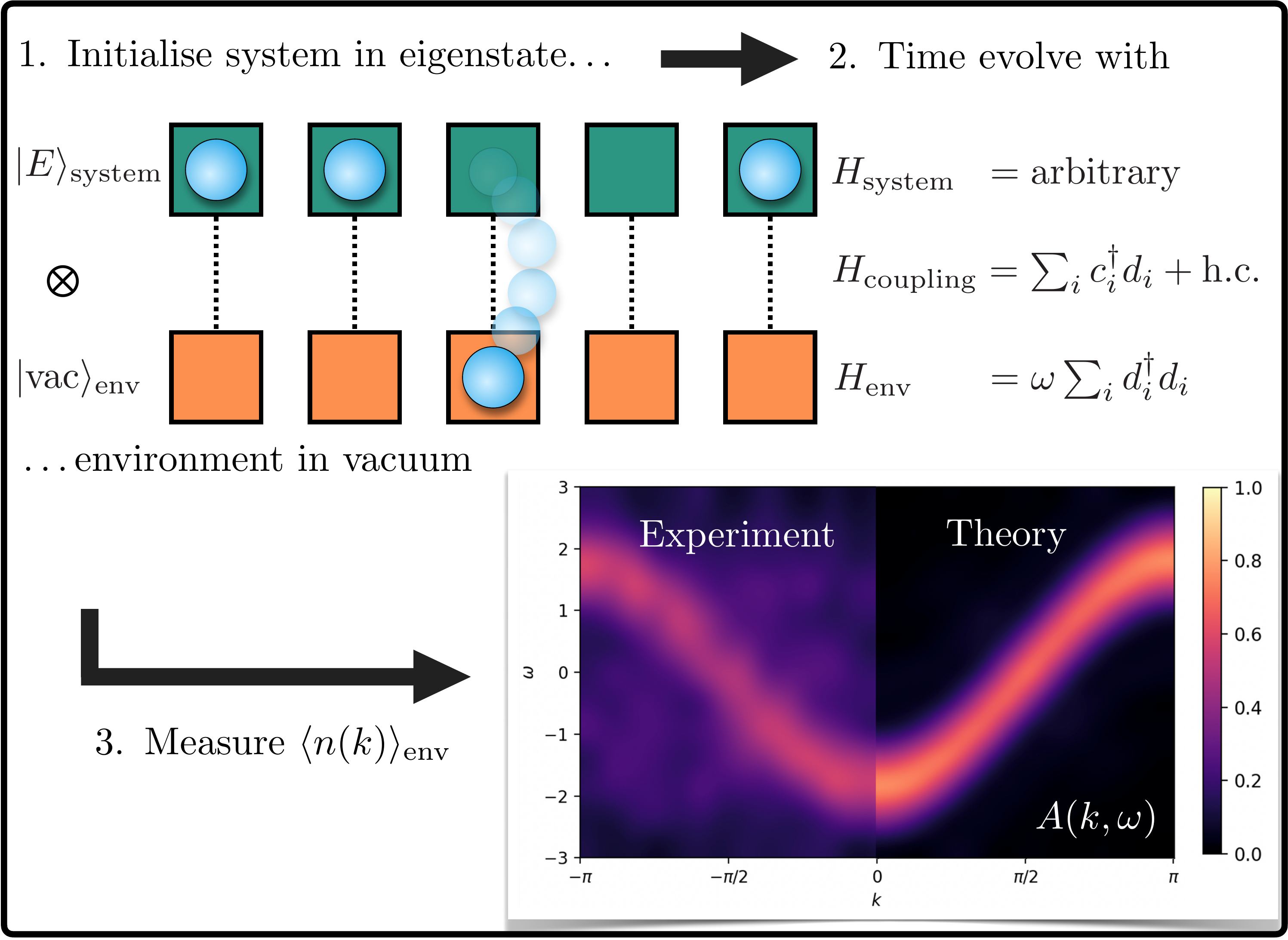}
    \caption{Schematic representation of our method for measuring spectral functions $A(k,\omega)$, and comparison of raw hardware results obtained from Quantinuum System Model H2 for $A(k,\omega)$ with noiseless expectation values of quantum circuits for a system of size $27$, using $54$ qubits. A bicubic interpolation is used for both experimental and exact data.}
    \label{fig:conceptual}
\end{figure}

Classically, spectral functions are thus typically computed through the dynamical correlations $\langle c^\dagger_j(t)c_0(0)\rangle$ and $\langle c_j(0)c^\dagger_0(t)\rangle$. To compute these complex amplitudes on quantum computers, one generally requires ancillas, linear-response theory approaches or Hadamard-test like approaches \cite{somma2002simulating,bauer2016hybrid,kanasugi2023computation,endo2020calculation,libbi2022effective,tazhigulov2022simulating,pedernales2014efficient,li2024utilizing,bishop2025quantum}. These schemes all suffer from one important shortcoming, namely that they involve measuring operators such as $c_j+c^\dagger_j$ that do not commute for different $j$, contrary to analogous bosonic quantities for neutron scattering experiments \cite{roggero2019dynamic,kokcu2024linear,francis2020quantum,chiesa2019quantum,kosugi2020linear,baez2020dynamical,sun2025probing}. Hence one can measure only one correlation at a time, resulting in a rather large shot overhead $\mathcal{O}(N)$ compared to settings where all $N$ measured bits contain useful information. Because of the cost of such experiments, only small system sizes have been implemented on actual quantum hardware \cite{libbi2022effective,greene2023quantum,kokcu2024linear,selisko2025dynamical,gomes2023computing,irmejs2025approximating,ehrlich2025variational,li2024utilizing,rizzo2022one}. This shot overhead is particularly problematic for ion-trap quantum computers, that despite their superior connectivity and two-qubit gate fidelity suffer from a relatively long shot time. There, a shot overhead $\mathcal{O}(N)$ could become impractical for utility scale. There exist other approaches to compute the spectral functions directly in frequency domain, but they typically require heavier linear algebra solving routines \cite{tong2021fast,chen2021variational,keen2021quantum,jamet2022quantum,jamet2021krylov,baker2021lanczos} or only access moments in frequency domain \cite{irmejs2025approximating}.

In this manuscript, we solve this problem by providing a way of directly measuring $A(k,\omega)$ on a quantum computer for all $N$ momenta $k$ at a fixed frequency $\omega$, using all $N$ measured bits of information. This is performed through an idealised modelisation of the ARPES experiment itself that is sketched in Fig \ref{fig:conceptual}. Namely, we show that by coupling the system to an environment with $N$ sites and using a fermionic Fourier transform (FFT), one can access $A(k,\omega)$ by just measuring on-site $Z$ Pauli matrices. Although requiring to double the number of qubits, and coming with a two-qubit gate overhead, the approach saves a factor $N$ in the number of shots and $\mathcal{O}(N)$ in overall runtime, drastically improving the scalability of the spectral function measurements, in particular on ion-trap quantum computers. Besides these theoretical developments, we test our method on Quantinuum's H2-2 ion-trap quantum computers on $54$ qubits, obtaining a faithful measured spectral function of a free-fermion Hamiltonian on $27$ sites. In order to be able to apply a FFT on $27$ sites, we used the recursive structure of FFT with radix $3$ and proposed an efficient implementation of the fermionic SWAPs required for generic radix.

\section{Spectral function through coupling with an environment}
\subsection{Description of the method}
We consider a system of $N$ sites with Hamiltonian $H_{\rm sys}$, and an environment with same number of sites $N$ with Hamiltonian $H_{\rm env}$. We denote by $c_j,c_j^\dagger$ for $j=1,...,N$ the fermionic operators on the system, and by $d_{j},d_j^\dagger$ those on the environment. The system and the environment are coupled through the total Hamiltonian
\begin{equation}\label{H}
    H=H_{\rm sys}+\epsilon H_{\rm int}+\omega H_{\rm env}\,,
\end{equation}
with $H_{\rm int}$ applying on both the system and the environment, and $\epsilon,\omega$ parameters that tune the strength of the different terms. We set the following
\begin{equation}
    H_{\rm env}=\sum_{j=1}^N d^\dagger_{j}d_{j}\,,\qquad H_{\rm int}=\frac{1}{2}\sum_{j=1}^N d^\dagger_{j}c_{j}+c^\dagger_{j}d_{j}\,,
\end{equation}
with the system Hamiltonian $H_{\rm sys}$ remaining arbitrary but particle-conserving. We initialize the total system at time $t=0$ in a tensor product of an eigenstate $|E\rangle$ of $H_{\rm sys}$ and of the vacuum state $|0\rangle$ for the environment. The system could also be initialized in a thermal state of $H_{\rm sys}$, or any diagonal ensemble in the energy basis. We then evolve the total system with $H$ for a fixed time $t$, and denote $\langle\cdot\rangle$ expectation values within the state $e^{iHt}|E\rangle \otimes |0\rangle$. At the end of the evolution, we measure the momentum occupation number $n(k)=d^\dagger(k)d(k)$ on the environment, with the Fourier transform $d(k)=\frac{1}{\sqrt{N}}\sum_{j=1}^N e^{-ijk}d_j$, where $k=2\pi n/N$ with $n=0,...,N-1$ an integer. Our result is that when $\epsilon\to 0$ we have
\begin{equation}\label{result}
    \langle n(k) \rangle=\epsilon^2 (A^+\star \hat{\varphi})(k,\omega)+\mathcal{O}(\epsilon^4)\,,
\end{equation}
with the function
\begin{equation}\label{kernel}
    \hat{\varphi}(\omega)=\frac{\sin^2(\omega t/2)}{\omega^2}\,,
\end{equation}
and where the convolution in frequency domain reads
\begin{equation}
    (A^+\star \hat{\varphi})(k,\omega)=\frac{1}{2\pi}\int_{-\infty}^\infty \D{\omega'}A^+(k,\omega')\hat{\varphi}(\omega-\omega')\,.
\end{equation}
In particular, when $t\to\infty$ we have $\hat{\varphi}(\omega)\sim t \delta(\omega)\pi/2 $ and so we obtain
\begin{equation}\label{limit}
   (A^+\star \hat{\varphi})(k,\omega)\underset{t\to\infty}{\sim}\frac{t}{4} A^+(k,\omega)\,.
\end{equation}
Let us now show the result \eqref{result}. We use time-dependent perturbation theory in $\epsilon$ to write
\begin{equation}\label{dyson}
\begin{aligned}
    &e^{itH}=e^{itH_0}+i\epsilon \int_{0}^t \D{s}e^{isH_0} H_{\rm int}e^{i(t-s)H_0}\\
    &-\epsilon^2 \int_{0}^t \D{s}\int_{0}^s \D{u}e^{iuH_0} H_{\rm int}e^{i(s-u)H_0}H_{\rm int}e^{i(t-s)H_0}+\mathcal{O}(\epsilon^3)\,,
\end{aligned}
\end{equation}
where we set $H_0= H_{\rm sys}+\omega H_{\rm env}$, and evaluate perturbatively in $\epsilon$ the expectation value $\langle 0|\otimes \langle E| e^{-iHt}n(k)e^{iHt}|E\rangle \otimes |0\rangle$. Since the mode occupation number on the environment $n(k)$ commutes with $H_0$, and since $n(k)|0\rangle=0$, the order $\epsilon^0$ of this expectation value vanishes. At order $\epsilon^1$, the application of only one $H_{\rm int}$ can only add a particle in the environment, and since it is not removed afterwards, this order also vanishes in the expectation value. One notes that this holds true for any odd power of $\epsilon$. Similarly, the only non-zero term at order $\epsilon^2$ comes from the order $\epsilon^1$ in $e^{\pm iHt}$ on both sides of $n(k)$ in the expectation value, because if one takes an order $\epsilon^2$ on one side and $\epsilon^0$ on the other side, one can commute $n(k)$ up to $|0\rangle$ on this other side, which vanishes. So we have
\begin{equation}
\begin{aligned}
    \langle n(k) \rangle=&\epsilon^2 \int_0^t \D{u}\int_0^t \D{s}
    \langle 0|\otimes \langle E| e^{-i(t-u)H_0}H_{\rm int} e^{-iu H_0}\\
    &\times n(k)e^{is H_0}H_{\rm int} e^{i(t-s)H_0}|E\rangle \otimes |0\rangle+\mathcal{O}(\epsilon^3)\,.
\end{aligned}
\end{equation}
Because $|E\rangle \otimes |0\rangle$ is an eigenstate of $H_0$ with energy $E$, we can write
\begin{equation}
\begin{aligned}
    &\langle n(k) \rangle=\epsilon^2 \int_0^t \D{u}\int_0^t \D{s}e^{iE(u-s)}\\
    &\langle 0|\otimes \langle E| H_{\rm int} e^{-iu H_0}
    n(k)e^{is H_0}H_{\rm int}|E\rangle \otimes |0\rangle+\mathcal{O}(\epsilon^3)\,.
\end{aligned}
\end{equation}
We write then the interaction Hamiltonian in momentum space $H_{\rm int}=\frac{1}{2}\sum_{q}d^\dagger(q)c(q)+c^\dagger(q)d(q)$. Using $d(q)|0\rangle=0$ and
\begin{equation}
    \langle 0|d(q) e^{-iu H_0}
    n(k)e^{is H_0} d^\dagger(q)|0\rangle=\delta_{k,q} e^{i\omega (s-u)}\,,
\end{equation}
we find
\begin{equation}
\begin{aligned}
    &\langle n(k) \rangle=\frac{\epsilon^2}{4} \int_0^t \D{u}\int_0^t \D{s}e^{i\omega(s-u)}\\
    & \langle E| e^{i(u-s) H_{\rm sys}}c^\dagger(k)e^{i(s-u) H_{\rm sys}}
    c(k)|E\rangle +\mathcal{O}(\epsilon^3)\,.
\end{aligned}
\end{equation}
Performing a change of variable $v=u-s$, $w=u+s$, this can be written
\begin{equation}
    \langle n(k) \rangle=\epsilon^2\int_{-\infty}^\infty \D{v} \varphi(v) e^{-i\omega v}\langle E| c^\dagger(k,v)c(k,0)|E\rangle +\mathcal{O}(\epsilon^3)\,,
\end{equation}
with $\varphi(v)=(t-|v|)/4$ for $|v|\leq t$, and $\varphi(v)=0$ for $|v|>t$. We introduced the time-evolved mode $c^\dagger(k,t)=e^{itH_{\rm sys}}c^\dagger(k)e^{-itH_{\rm sys}}$. This expression can be seen as the Fourier transform of $\varphi(v)$ times $\langle E| c^\dagger(k,v)c(k,0)|E\rangle$. The Fourier transform  of $\varphi(v)$ is $\hat{\varphi}(\omega)$ in \eqref{kernel}, and the Fourier transform in $v$ of $\langle E| c^\dagger(k,v)c(k,0)|E\rangle$ is precisely $A^+(k,\omega)$. Using that the Fourier transform of the product of two functions is the convolution of their Fourier transforms divided by $2\pi$, we thus get our result \eqref{result}.

The full spectral function $A(k,\omega)$ also involves $\langle E| c(k,0)c^\dagger(k,v)|E\rangle$. By repeating analogous calculations, one can check that this can be obtained by initially preparing the environment in the completely filled state $|{\rm all}\rangle$ where there is one fermion on every site, instead of the vacuum $|0\rangle$. In that case we obtain
\begin{equation}\label{result2}
    \langle 1-n(k) \rangle=\epsilon^2 (A^-\star \hat{\varphi})(k,\omega)+\mathcal{O}(\epsilon^4)\,.
\end{equation}
The spectral function $A(k,\omega)$ is thus obtained as the sum of these two settings.\\

Some comments on the method are in order. Firstly, the method gives access to the convolution of the spectral function $A(k,\omega)$ with a kernel $\hat{\varphi}$ in \eqref{kernel}. Only in the limit $t\to\infty$ in \eqref{limit} do we recover $A(k,\omega)$ without convolution. For finite time evolution $t$, the effect of this convolution is to blur the resulting spectral function in the frequency direction and to broaden bands with a width $\sim 1/t$. A similar effect would occur by computing the dynamical correlations $\langle c^\dagger(k,t)c(k,0)\rangle$ in \eqref{aa} and taking the Fourier transform over a finite time window and is thus not specific to our method. Moreover, we note that even in experimental outcomes of ARPES, bands are always broadened in $\omega$ due to finite time windows as well \cite{iwasawa2020high}. 

The influence of the coupling $\epsilon$ on the computation of the spectral function is however specific to our computation method. In principle, to recover the spectral function from \eqref{result}, one has to divide the expectation value of $n(k)$ by $\epsilon^2$ and take $\epsilon \to 0$. On a quantum computer, this amplifies shot noise by $1/\epsilon^2$, and thus multiplies the number of shots by $1/\epsilon^4$ to get a same precision. It becomes thus prohibitively expensive to take the limit $\epsilon \to 0$. In practice we suggest fixing a finite $\epsilon$ and avoiding any division by $\epsilon$. This is justified by the fact that overall constants are typically irrelevant in photoemission spectroscopy. However, it is crucial to understand whether there are any qualitative consequences of taking a finite coupling $\epsilon = \mathcal{O}(1)$ beyond the leading order \eqref{result}. 

Next, we note that the algorithm requires to measure the mode occupation number in momentum space in the environment $n(k)=d^\dagger(k)d(k)$. This quantity is not directly accessible after time-evolving with $H$, because the measurement of each environment site would yield $d_j^\dagger d_j$ the density in real space and its correlations. To measure $n(k)$, one must perform a fermionic Fourier transform (FFT) on the environment at the end of the time evolution. This will be detailed in another section below. Then $n(k)$ is obtained by a simple measurement of $Z$ on all sites.

Finally, let us mention that there is some freedom to choose other Hamiltonians on the environment. If one considers a more general free-fermionic Hamiltonian $H_{\rm env}=\sum_k \varepsilon_k d^\dagger(k) d(k)$ in Fourier space, for some dispersion relation $\varepsilon_k$, then one can see that one measures $A(k,\varepsilon_k)$ for all values of $k$ with a $k$-dependent $\omega=\varepsilon_k$. This can be used to scan a non-trivial curve in one shot in the $k,\omega$ plane, instead of horizontal lines at fixed $\omega$. However, these environment Hamiltonians with non-flat bands would come with an extra two-qubit gate overhead.

\subsection{Effect of finite coupling $\epsilon$} To understand the effect of finite $\epsilon$, the example of a free fermion Hamiltonian
\begin{equation}
    H_{\rm sys}=\nu\sum_{j=1}^N c_j^\dagger c_{j+1}+c_{j+1}^\dagger c_{j}
\end{equation}
 with some parameter $\nu$, is insightful, because exact calculations can be performed for finite $\epsilon$. We find in our protocol with an empty environment (see Supplemental Material)
\begin{equation}\label{ffexact}
    \langle n(k) \rangle=\frac{\epsilon^2\sin^2(t\Omega)}{\epsilon^2+\left(\omega-2\nu\cos k\right)^2}\rho_k\,,
\end{equation}
with $\rho(k)=\langle E| c^\dagger(k)c(k) |E\rangle$ evaluated in the initial state, and with
\begin{equation}
    \Omega=\frac{1}{2}\sqrt{\epsilon^2+\left(\omega-2\nu\cos k\right)^2}\,.
\end{equation} 
At leading order in $\epsilon$, this is
\begin{equation}\label{ffexpand}
    \langle n(k) \rangle=\epsilon^2\frac{\sin^2[(\omega-2\nu\cos k)t/2]}{\left(\omega-2\nu\cos k\right)^2}\rho_k+\mathcal{O}(\epsilon^4)\,.
\end{equation}
In this free fermion case we have $\langle c^\dagger(k,t)c(k,0)\rangle=e^{2it\nu \cos k}\rho_k$, so we get $A^+(k,\omega)=2\pi\delta(\omega-2\nu\cos k) \rho_k$. Eq. \eqref{ffexpand} agrees thus with the general formula \eqref{result}.

Now, in this free fermion case, \eqref{ffexact} also contains all next orders in $\epsilon$. We observe that here, all orders in $\epsilon$ are obtained by replacing $\hat{\varphi}$ in \eqref{result} by
\begin{equation}
\hat{\varphi}_\epsilon(\omega)=\frac{\sin^2(t\sqrt{\epsilon^2+\omega^2}/2)}{\epsilon^2+\omega^2}\,.
\end{equation}
Let us study the consequences of finite $\epsilon$ in \eqref{ffexact}. When $\epsilon\to 0, t\to\infty$, all the weight is localized at $\omega=2\nu \cos k$. For finite $\epsilon,t$, the sinus in \eqref{ffexact} vanishes a first time away from $\omega=2\nu\cos k$ when $\frac{t}{2}\sqrt{\epsilon^2+\left(\omega-2\nu\cos k\right)^2}=n\pi$, with $n$ the integer such that $(n-1)\pi \leq\epsilon t/2<n\pi$. This gives a band broadening in the $\omega$ direction of
\begin{equation}
    \Delta \omega=\frac{2n\pi}{t}\sqrt{1-\left(\frac{\epsilon t}{2n\pi}\right)^2}\,.
\end{equation}
Even when $\epsilon=0$, there is a broadening in the $\omega$ direction due to finite integration time. Finite coupling $\epsilon$ actually contributes to decreasing this broadening when $\epsilon t<2\pi$.

\begin{figure}
\begin{picture}(100,100)
\put(-73,0){\includegraphics[width=\linewidth]{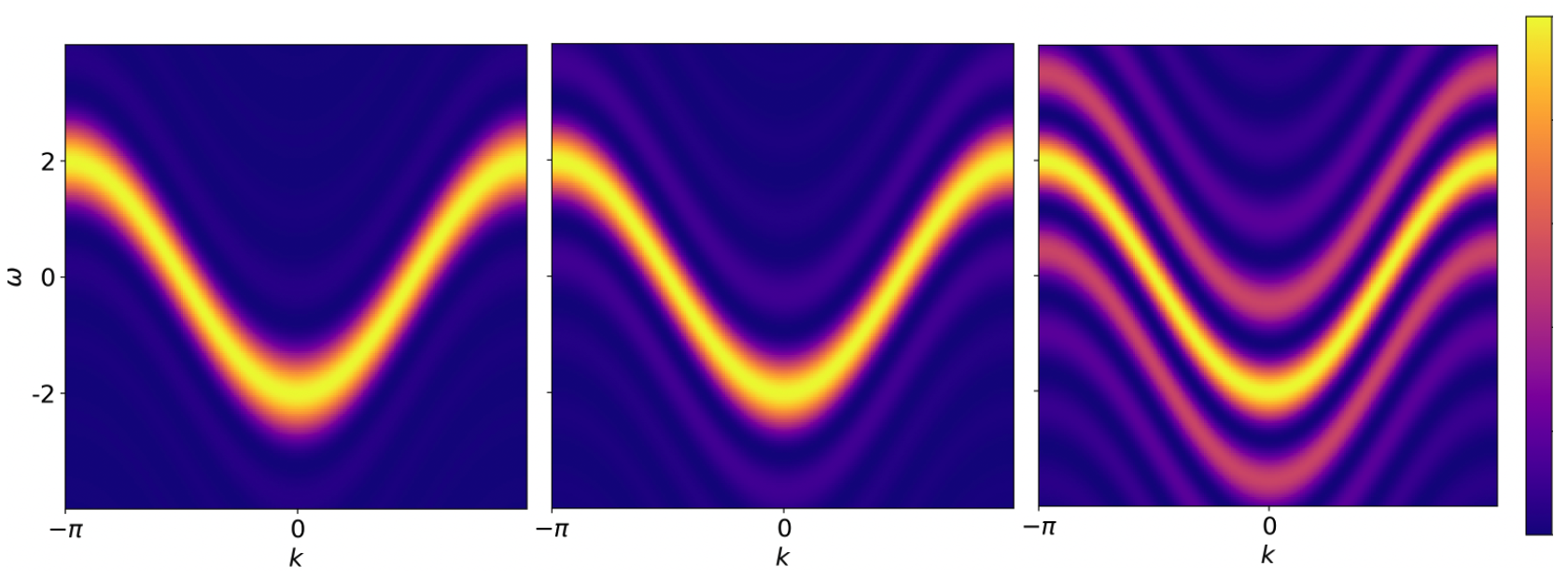}}
\put(-18,78){\color{white}\tiny $\epsilon=0.01$}
\put(60,78){\color{white}\tiny $\epsilon=\pi/t$}
\put(127,78){\color{white}\tiny $\epsilon=1.5\pi/t$}
\end{picture}
    \caption{Amplitude $\langle n(k)\rangle$ for $H_{\rm sys}$ a free-fermion hopping Hamiltonian on $200$ sites, as a function of $k$ and $\omega$, computed with $t=5$, for different values of $\epsilon$. The three panels have a different color scale, ranging from value zero (darker) to the maximal value (lighter).}
    \label{fig:spectral_ff}
\end{figure}

Beyond this main band broadening, finite $t$ or $\epsilon$ also leads to the appearance of ghost bands. The first one is obtained approximately for $\frac{t}{2}\sqrt{\epsilon^2+\left(\omega-2\nu\cos k\right)^2}=n\pi+\frac{t\epsilon}{2}$. The ratio of amplitude between the next band and the main band is then approximately
\begin{equation}
    r=\left(\frac{t\epsilon }{2n\pi+t\epsilon }\right)^2\,.
\end{equation}
Taking $t\epsilon \leq \pi$ ensures that $r\leq 1/9$ and the next bands have a relatively negligible amplitude. These different effects are shown in Fig \ref{fig:spectral_ff}, where we display color plots of \eqref{ffexact} for different values of $\epsilon$ at finite $t$. At fixed $t$, one sees that the ghost bands are negligible up to $\epsilon=\pi/t$, while for larger values of $\epsilon$ they become strongly visible.

These insights from the free fermion model can be confirmed and complemented by a strong coupling expansion instead of a small $\epsilon$ expansion. This is equivalently obtained by putting a coefficient $\nu$ in front of $H_{\rm sys}$ in \eqref{H} and doing an expansion at small $\nu$ and fixed $\epsilon$. At leading order in $\nu$ we find for any system Hamiltonian $H_{\rm sys}$
\begin{equation}
    \begin{aligned}
       & \langle n_k\rangle=\sin^2(t\Omega_0)\frac{\epsilon^2}{\omega^2+\epsilon^2} \langle E| c^\dagger(k) c(k) |E\rangle+\mathcal{O}(\nu)\,,
    \end{aligned}
\end{equation}
with $\Omega_0=\frac{1}{2}\sqrt{\omega^2+\epsilon^2}$ (see Supplemental Material). The quantity $\langle E| c^\dagger(k) c(k) |E\rangle$ is the integral of $A(k,\omega)$ over $\omega$, corresponding to no time evolution in the system. Notably, we see the same renormalization of the frequency $\omega$ into $\sqrt{\epsilon^2+\omega^2}$ as in the free fermion case.

From these different insights we can conclude that the effect of a finite coupling $\epsilon$ is expected to be not much visible when $\epsilon t\lessapprox \pi$. This can be understood physically by noting that for a coupling $\epsilon$, the time that it takes to bring a particle from the system to the environment is $t \sim 1/\epsilon$. So when $\epsilon t \ll \pi$, particles do not have time to go back and forth several times between the system and the environment, which approximates more faithfully the spectral function, corresponding to just one movement of a particle from the system to the environment. We note that higher orders in $\epsilon$ that come from simultaneous excitation of multiple particles across the system are more physical, because also present in actual ARPES experiments. Only the reintroduction of a particle into the system after it has left it is unphysical.

\subsection{Comparison with dynamical correlations approach} Let us compare our method with the more standard approach consisting in measuring the dynamical correlations $\langle c_j^\dagger(t)c_0(0)\rangle$ for several times $t$ and sites $j$, and then performing a Fourier transform to obtain the spectral functions in \eqref{aa}. This is typically done with variations of the following scheme. One creates a superposition state by applying $e^{i\theta c_0}$ on the prepared eigenstate $|E\rangle$ for some angle $\theta$. Assuming that such on odd fermionic operator is available in a given fermionic encoding, one then performs a time evolution with $H$ for time $t$, and then one measures $c_j+c_j^\dagger$. A fundamental limitation of this approach is that since different $c_j+c_j^\dagger$ do not commute, one can measure only one at a time, requiring to repeat $N$ shots to obtain all values of $j$ and compute the Fourier transform of \eqref{aa} in the space direction. With our method, one sampling value for $A^\pm(k,\omega)$ for all values of $k$ are obtained directly in one shot. There is thus a reduction by $\mathcal{O}(N)$ of the number of shots required to obtain a given precision on the result. Moreover, with our method the Fourier transform in the time domain is performed automatically on hardware. This way, specific values of $\omega$ can be targeted without having to compute a large number of time points $t$, as would be required with an approach through dynamical correlations. It is indeed the case that one is often interested in values of frequencies near the Fermi level. This entails thus another significant shot reduction. 

Nevertheless, the shot reduction of our method comes with a qubit overhead, namely one requires twice more qubits because of the need to include the environment in the simulation. It also comes with a two-qubit gate overhead, that is $2N$ extra two-qubit gates to perform per Trotter step, and one single extra fermionic Fourier transform, which can be carried out with $\mathcal{O}(N \log^2 N)$ two-qubit gates \cite{constantinides2025low}. For local Hamiltonians $H$, this will entail an overhead $\mathcal{O}(1)$ in terms of two-qubit gates per Trotter step. Despite these overheads, the algorithm we presented displays a significant shot reduction that can have a fundamental impact on the practicality of the computation of spectral functions on quantum computers as systems of larger sizes are studied. This is particularly true for ion-trap quantum computers. These platforms have significantly lower two-qubit gate infidelity than other platforms and enjoy effective all-to-all connectivity. They are however slower, which sometimes makes the shot count the actual bottleneck of quantum computations. In this perspective, a reduction of $N$ in the number of shots, especially for utility scale hardware where we envision $N$ to be several hundreds or thousands, is a fundamental improvement in the scalability of quantum computing spectral functions.

As a concrete example, let us consider the experiment that we ran on Quantinuum's H2-2 hardware, detailed below. Focusing on the free fermion experiment, we had $2172$ two-qubit gates per circuit. The overhead due to our method (namely, all the two-qubit gates that apply on environment qubits) is $1298$. Despite this two-qubit gate overhead, we spare a factor $\sim 27$ in the number of shots to get the same precision. To obtain the same color plot of the spectral function without our method, the total cost (in terms of runtime, or total number of two-qubit gates implemented) would have been around $\sim 11$ times more expensive, and therefore probably difficult to implement. This cost difference scales with system size and would be even larger for larger systems or two-dimensional systems.

\begin{figure}
\begin{picture}(100,120)
\put(-73,0){
\includegraphics[width=\linewidth]{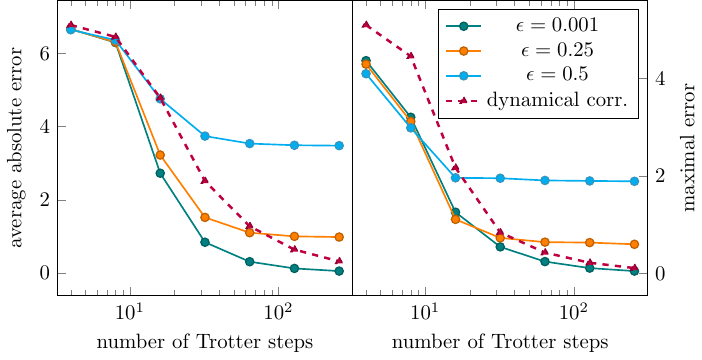}
}
\put(-60,13){\color{black}\textbf{(a)}}
\put(45,13){\color{black}\textbf{(b)}}
\end{picture}
    
    \caption{Average absolute error (a) and maximal error (b) between spectral function computed with perfect time evolution, and computed  with a finite number of Trotter steps, comparing our environment method with different $\epsilon$'s, and the standard dynamical correlation approach, for the Hamiltonian \eqref{Hamiltonian} with an additional interaction term $V\sum_{j=1}^N c^\dagger_j c_j c^\dagger_{j+1}c_{j+1}$ with $V=4$, for a total evolution time $t=5$, in size $N=9$, for $26$ values of $\omega$. For each point an overall scaling factor for all values of $k,\omega$ is optimized so as to minimize the difference with exact.}
    \label{fig:trotter}
\end{figure}

Let us finally comment on the effect of imperfect time evolution and finite Trotter step size on the spectral function obtained. The equivalence between our approach \eqref{result} and the spectral function \eqref{aa} crucially relies on the fact that the state prepared $|E\rangle$ is an eigenstate of the time evolution operator $e^{itH}$. If this is not the case, for example because of Trotter error when implementing $e^{itH}$ on a digital quantum computer, then \eqref{result} and \eqref{aa} will differ. However, one can argue that our expression \eqref{result} will be more accurate than \eqref{aa}, because it is closer to what is actually measured in experiments. In fact, some properties of the spectral function as computed in \eqref{result} will still remain true with an imperfect time evolution, whereas they will break down in \eqref{aa}. For example, it is well-known that the spectral function is a positive quantity, as made evident by its Lehmann representation
\begin{equation}
    A^+(k,\omega)=2\pi \sum_{E_n} |\langle E|c^\dagger(k)|E_n\rangle|^2 \delta(\omega-E+E_n)\,.
\end{equation}
However, if there is Trotter error in the time evolution in \eqref{aa}, then the spectral function as computed in \eqref{aa} can be negative. On the other hand, the quantity \eqref{result} is always positive, because it is the expectation value of $n(k)=c^\dagger(k)c(k)$ which takes value $0$ or $1$, and that is $0$ when $\epsilon=0$. This holds true even if the state prepared $|E\rangle$ is not an exact eigenstate of the time evolution operator due to Trotter error. To illustrate this lower sensitivity to Trotter error of our approach, we plot in Fig \ref{fig:trotter} the error obtained on the spectral function $A(k,\omega)$ as a function of number of Trotter steps at a fixed total evolution time, comparing our approach with that consisting in taking the Fourier transform of the dynamical correlations, for $N\times 26$ values of $k,\omega$. For $\epsilon\to 0$, we observe indeed that our approach has significantly less Trotter error, achieving the same precision with a number of Trotter steps a multiple times smaller. For finite $\epsilon$, there is a finite discrepancy even for continuous-time evolution. However, we observe that at small to intermediate number of Trotter steps, the overall error is smaller than for the dynamical correlation approach.

\begin{figure*}
\begin{picture}(100,160)
\put(-200,20){
\includegraphics[width=0.45\linewidth]{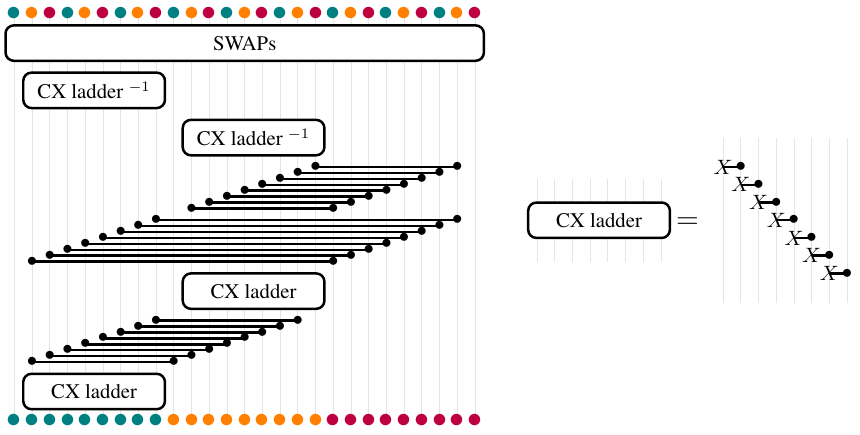}
$\qquad$
\includegraphics[width=0.24\linewidth]{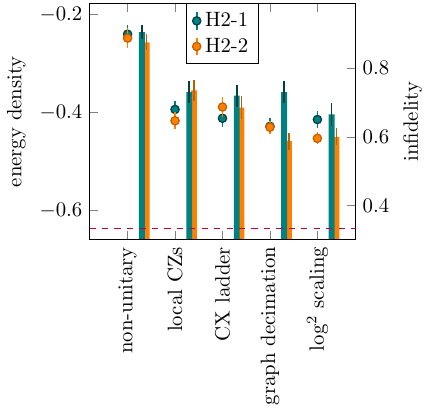}
$\quad$
\includegraphics[width=0.24\linewidth]{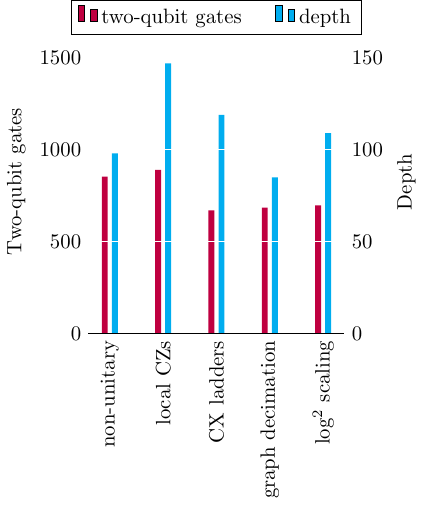}
}
\put(-200,10){\color{black}\textbf{(a)}}
\put(50,10){\color{black}\textbf{(b)}}
\put(185,10){\color{black}\textbf{(c)}}
\end{picture}

    \caption{\emph{Fast Fermionic Fourier Transform.}  
    \textbf{(a):} Circuit implementing on $27=3^3$ qubits a $3$-way interleave operation (from bottom to top) or inverse interleave operation (from top to bottom). The colored dots at the top and bottom indicate how the qubits are permuted. \textbf{(b):} Energy density (dots) and infidelity (bars) of a free-fermion ground state on $27$ sites, measured on hardware, for the machines H2-1 and H2-2, and for different implementations of the interleaves (see text). The dotted red line indicates the ground state energy density.  \textbf{(c):} Two-qubit gate count and two-qubit gate depth of the circuits run on hardware appearing in panel (b).}
    \label{fig:fermionic_swaps}
\end{figure*}

\section{Fermionic Fourier transform}
\subsection{Generic radix} 
\begin{figure}
\includegraphics[width=0.84\linewidth]{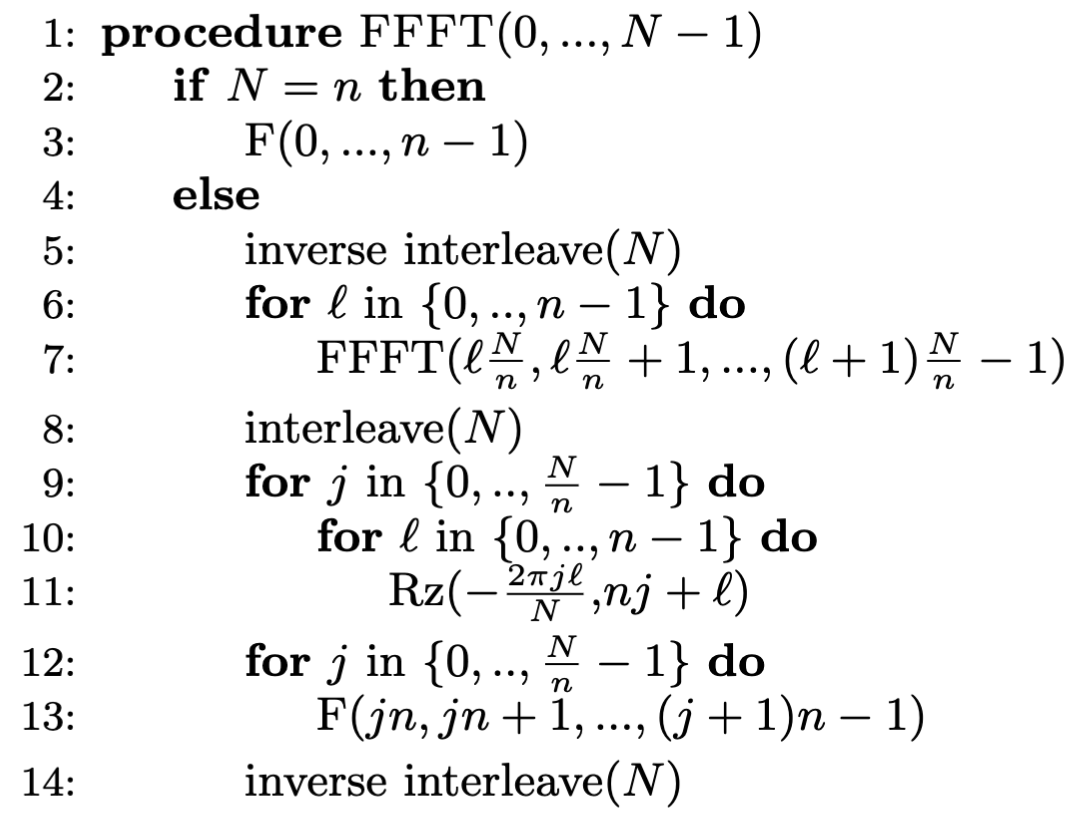}
    \caption{Recursive algorithm for the FFT on $N=n^k$ modes. $Rz(\alpha,i)$ denotes a rotation of angle $\alpha$ in the $Z$ direction on qubit $i$, namely the operator $e^{-iZ_i \alpha/2}$, and $F$ denotes the FFT on $n$ modes.}\label{algo_1}
\end{figure}

Our scheme requires the implementation of a FFT. This is defined by the unitary operator $F_N$ that maps $c_j$ on $N$ modes to $c(k)=F_N^\dagger c_j F_N$ in momentum space with $k=2\pi j/N$ \cite{verstraete2009quantum}. Namely, $F_N$ is the operator such that for all $j=0,...,N-1$
\begin{equation}
    F_N^\dagger c_j F_N=\frac{1}{\sqrt{N}}\sum_{\ell=0}^{N-1} e^{-2i\pi \ell j/N}c_\ell\,.
\end{equation}
This definition in matrix notation means that a quantum circuit that implements $F_N$ maps in the Heisenberg picture the operator $c_j$ to $c(k)$. For small system sizes, efficient circuits can be found. We give here expressions that are written in \emph{circuit order} for $N=2$ and $N=3$, namely first symbol applies first. For $N=2$ we have
\begin{equation}
   F_2=e^{i(-Y_1X_2+X_1Y_2)\pi/8} Z_2\,,
\end{equation}
which uses two two-qubit gates, and for $N=3$
\begin{equation}
\begin{aligned}
F_3=&e^{i(-Y_2X_3+X_2Y_3)\pi/8}   e^{i(-Y_1X_2+X_1Y_2)\beta/2}\\
    &\quad \times   e^{-i(X_2X_3+Y_2Y_3)\pi/8}  Z_2 S_3^\dagger\,,
\end{aligned}
\end{equation}
with $\cos\beta=1/\sqrt{3}$, $\sin\beta=\sqrt{2/3}$,
which uses six two-qubit gates. For larger system sizes, it is well-known that the FFT has a recursive structure, like its classical counterpart, called the fast fermionic Fourier transform (FFFT). This recursive structure for FFT is always presented in the literature for system sizes of the form $N=2^k$ for $k$ an integer \cite{verstraete2009quantum,campbell2022early}. However, it also applies to other radix different than $2$, namely to any system size of the form $N=n^k$ for $n$ and $k$ two integers. Although it was developed for the non-fermionic quantum Fourier transform \cite{camps2021quantum}, it does not seem to have been developed for FFT. More generally, it can be applied to any prime factor decomposition of $N$. This recursive structure is based on the following equation
\begin{equation}\label{recusrion}
\begin{aligned}
    &F_N^\dagger c_j F_N=\\
    &\frac{1}{\sqrt{n}}\sum_{\ell=0}^{n-1}e^{-2i\pi a\ell/n}\frac{e^{-2i\pi b\ell/N}}{\sqrt{N/n}}\sum_{q=0}^{N/n-1}e^{-2i\pi bq/(N/n)}c_{nq+\ell}\,,
\end{aligned}
\end{equation}
where we decomposed $j=aN/n+b$, with $a=0,...,n-1$ and $b=0,...,N/n-1$. The $n$ sums over $q$ are FFTs on system sizes $N/n$. However, contrary to classical or non-fermionic Fourier transforms, the recursion cannot be directly applied at this stage, because these FFT differ from the original FFT on $N$ modes for the following reason. These FFTs apply on the modes $c_{nq+\ell}$ for $q=0,...,N/n-1$ at fixed $\ell$: When using a Jordan-Wigner transformation to implement the fermionic encoding of $c_j$ into qubits, taking as JW order $c_0,c_1,...,c_{N-1}$, the modes $c_{nq+\ell}$ are not adjacent, so that fermionic operations on these modes will not be translated into the same qubit operations as for $F_N$. To make the recursion explicit, one must swap the fermionic modes in such a way that they are adjacent for these FFTs on $N/n$ sites. Since the modes are fermionic, these swaps must be ``fermionic swaps". This corresponds to adding a ${\rm CZ}_{j,j+1}$ operation whenever two adjacent modes $j,j+1$ are swapped. To make all the modes $c_{nq+\ell}$ for $q=0,...,N/n-1$ adjacent, one must apply a so-called \emph{inverse interleave} operation that fermionic-swaps modes $0,1,...,N-1$ into $0,n,2n,...,1,n+1,2n+1,...,2,n+2,2n+2,...$, namely first all the $N/n$ integers equal to $0$ modulo $n$ by increasing order, then all the $N/n$ integers equal to $1$ modulo $n$ by increasing order, etc. The inverse operation, the interleave operation, fermionic-swaps $0,1,...,N-1$ into $0,N/n,2N/n,...,(n-1)N/n,1,1+N/n,1+2N/n,...$. Then, the sum over $\ell$ in \eqref{recusrion} is itself a FFT on $n$ modes, applied on the output of the recursive FFTs on $N/n$ modes, multiplied by the so-called twiddle factors $e^{-2i\pi b\ell/N}$. For completeness,  we write in Fig.~\ref{algo_1} an explicit algorithm for arbitrary radix-$n$ FFFT, namely that applies to system sizes of the form $N=n^k$ with $n,k$ some integers. It takes and returns the real-space and momentum-space fermionic modes in natural order. We note that if the FFFT is applied on a product state in the $Z$ basis, or if it is applied at the end of the circuit before measuring in the $Z$ basis, then the initial or final inverse interleave operations can be changed into a simple qubit relabeling, without any quantum operations.

\begin{figure}
\includegraphics[width=\linewidth]{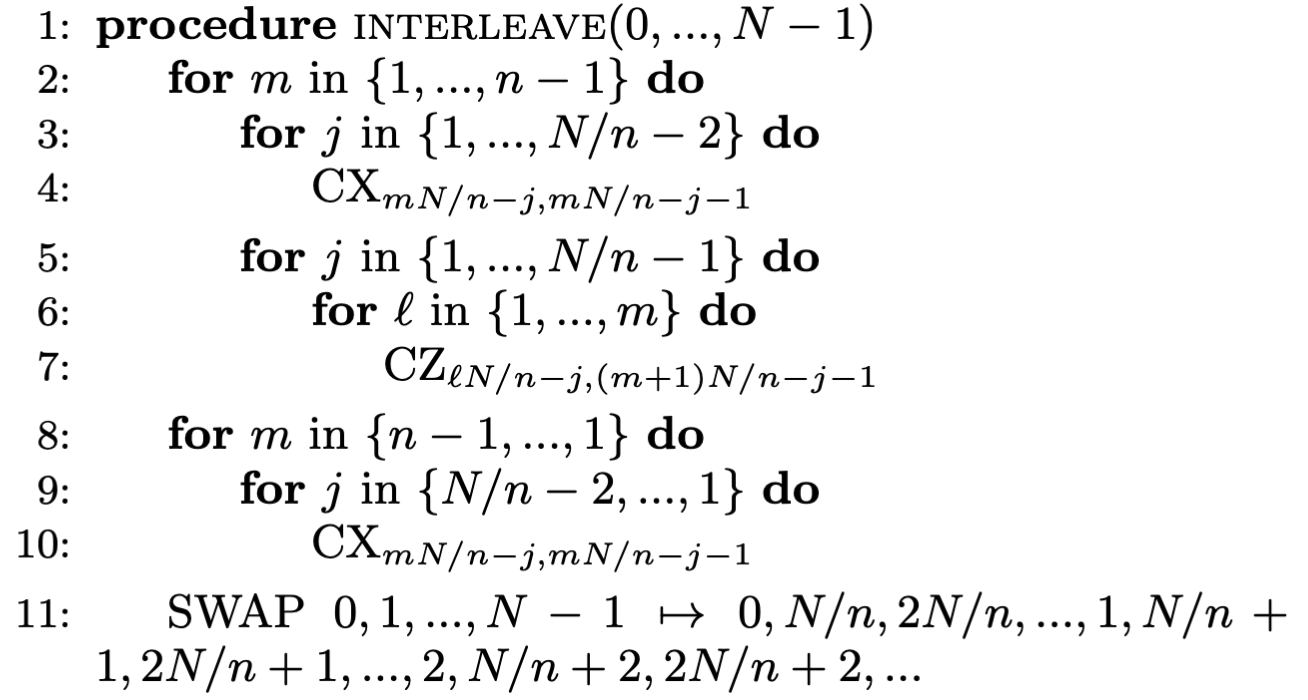}
            
    \caption{Interleave operation with radix $n$ on $N$ modes, for all-to-all coupled architectures where the SWAPs can be performed by relabeling qubits.}\label{algo_2}
\end{figure}

The FFFT on these $N$ modes requires two elementary functions: the FFT on $n$ fermionic modes (denoted $F$ in Fig.~\ref{algo_1}) 
, and the interleave operation. 
The depth scaling of the FFFT with $N$ is $\mathcal{O}(I_N\log_n N)$ where $I_N$ is the depth of one interleave operation. This interleave operation can be implemented in different ways. The simplest way is to decompose the permutation $\sigma$ that it implements into local transpositions $(j,j+1)$ that permute only neighbouring modes $j$ and $j+1$, namely to write $\sigma=(j_1,j_1+1)(j_2,j_2+1)...(j_m,j_m+1)$ with some sequence of integers $j_1,...,j_m$. Then $I_N$ can be written as
\begin{equation}
    I_N={\rm FSWAP}_{j_m,j_m+1}...{\rm FSWAP}_{j_1,j_1+1}\,,
\end{equation}
with the fermionic swap ${\rm FSWAP}_{j,j'}$ being the product of ${\rm SWAP}_{j,j'}$ and ${\rm CZ}_{j,j'}$. This decomposition requires $\mathcal{O}(N^2)$ gates and has depth $\mathcal{O}(N)$. It is mostly adapted to architectures with only local connectivity between qubits. On all-to-all coupled devices, it is advantageous to perform all the SWAPs at software level, and to optimize the application of the remaining CZs. In Ref. \cite{maskara2025fast} was provided an efficient re-organization of this interleave operation for radix $n=2$, using ladders of CXs and sequences of CZs that can be completely parallelized. We propose a generalization of this procedure for arbitrary radix $n$ that is adapted to all-to-all coupled devices where SWAP operations can be done implicitly in software. We explicitly write out this algorithm in Fig. \ref{algo_2}. We also illustrate it in panel a of Fig.~\ref{fig:fermionic_swaps} in the case of radix $n=3$ for $N=27=3^3$.

Such an implementation of the interleave operation reduces the number of gates to $\mathcal{O}(N)$, but has still depth $\mathcal{O}(N)$. In fact, we observe that this decomposition into CX ladders has minimal two-qubit gate count among the different methods tested for $N=27$. To reduce the depth, each of these CX ladders (appearing in panel a of Fig \ref{fig:fermionic_swaps}) can be further reorganized into depth $\mathcal{O}(\log N)$ with a unitary circuit \cite{constantinides2025low,remaud2025ancilla}, and even depth $\mathcal{O}(1)$ if one uses ancillas and measurement with feed-forward \cite{maskara2025fast,raussendorf2003measurement,van2004graphical}. Finally, since the interleave is a CZ circuit, one can optimize its decomposition into Clifford gates using the graph decimation technique \cite{doherty2026fast}. While this technique was originally describe to prepare graph states, it can be straightforwardly modified to implement CZ circuits, as described in Supplemental Material. We find that this approach gives gate sequences that have overall minimal depth, even compared to the non-unitary compilation, while retaining low two-qubit gate count (see Supplemental Material for optimal gate sequences).

\begin{figure}

\begin{picture}(100,125)
\put(-73,0){
\includegraphics[width=\linewidth]{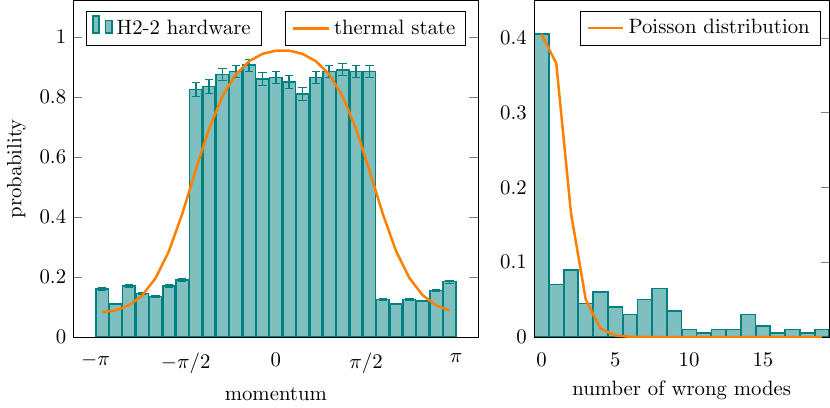}
}
\put(-60,5){\color{black}\textbf{(a)}}
\put(70,5){\color{black}\textbf{(b)}}
\end{picture}
    \caption{\textbf{(a)}: probability of measuring a momentum mode on H2-2 hardware after state preparation with the $\log^2(N)$ scaling compilation (noiseless is $1$ for momenta between $-\pi/2$ and $\pi/2$, and $0$ for the remaining modes), comparing with a thermal state with same energy density. \textbf{(b)}: probability of measuring a certain number of wrong modes, compared with a Poisson distribution with same probability of zero wrong mode measured.}
    \label{fig:histogram_fft}
\end{figure}

\begin{figure*}
\begin{picture}(100,120)
\put(-200,10){
\includegraphics[width=0.45\linewidth]{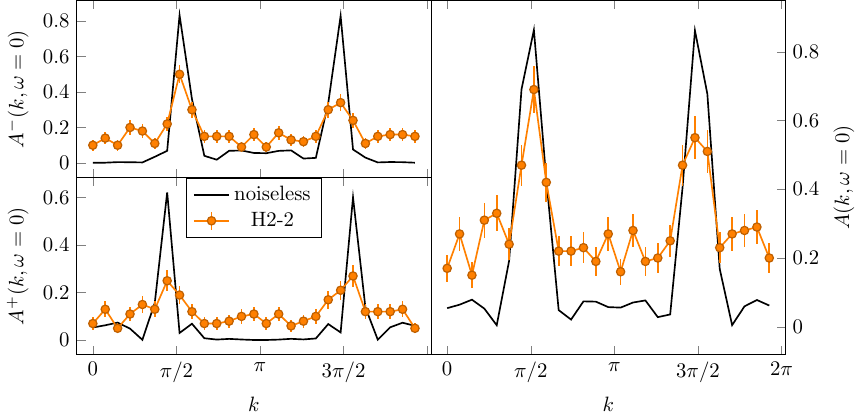}
\includegraphics[width=0.237\linewidth]{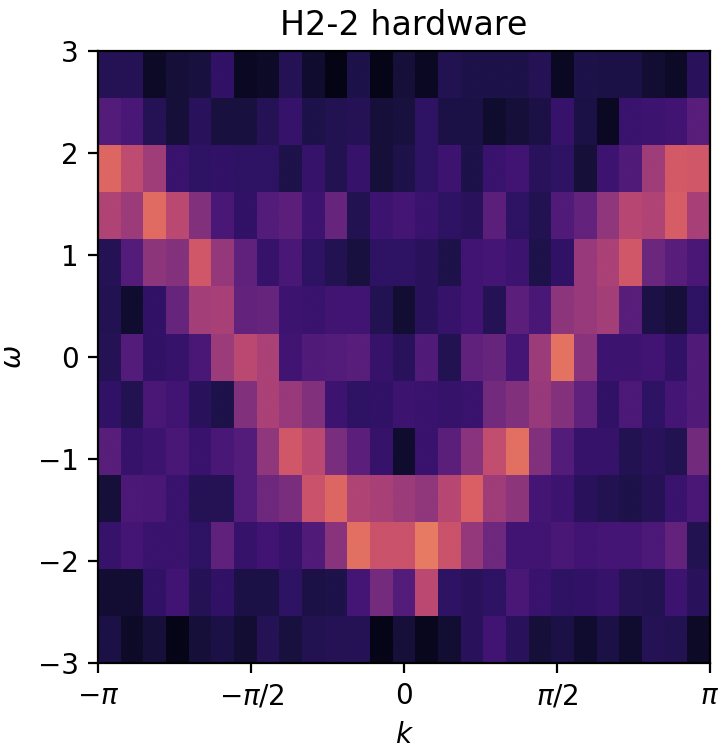}
\includegraphics[width=0.271\linewidth]{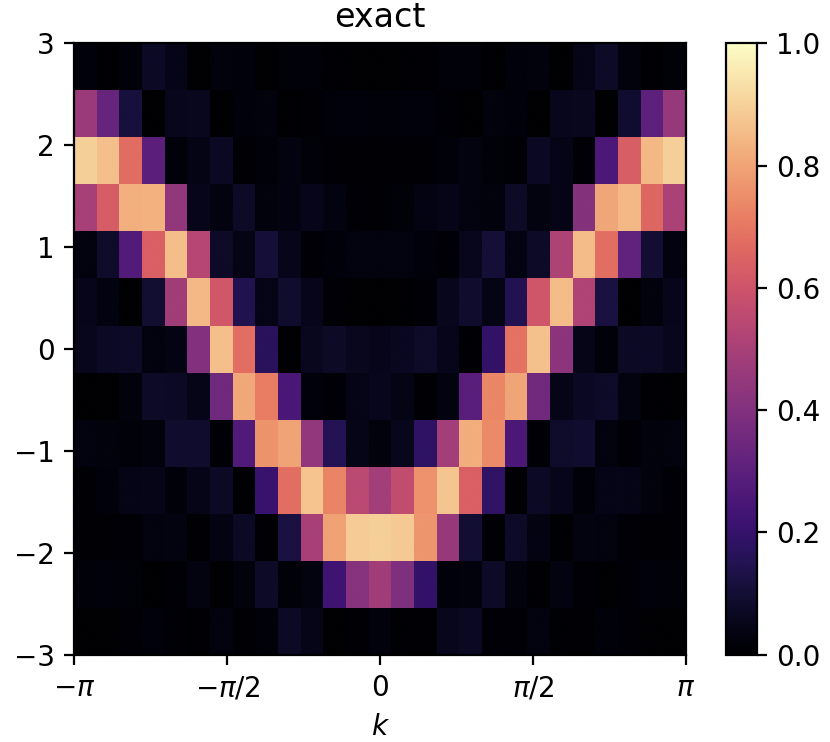}
}
\put(-195,10){\color{black}\textbf{(a)}}
\put(30,10){\color{black}\textbf{(b)}}
\put(155,10){\color{black}\textbf{(c)}}
\end{picture}

    \caption{\emph{Hardware results.} \textbf{(a):} Spectral functions $A^\pm(k,\omega)$ and $A(k,\omega)$ as a function of $k$ for $\omega=0$, for the $27$-site free fermion Hamiltonian described in the text, comparing hardware experiment and noiseless expectation values, evaluated with our method with $\epsilon=0.5$ and $t=5$. \textbf{(b, c):} Color plot for $A(k,\omega)$ comparing raw hardware data with exact noiseless values. 
    }
    \label{fig:hardware}
\end{figure*}

\subsection{Benchmark on hardware} 
We tested these five implementations of the FFFT on $27$ modes on hardware. We first prepare the ground state of a free-fermion Hamiltonian on $N=27$ sites using a FFFT. In momentum space, this Hamiltonian reads $H_{\rm sys}=\sum_{k}\epsilon_k c^\dagger(k)c(k) $ with $\epsilon_k=-2\cos k$, and its ground state is obtained by a product state in momentum space where every momenta $k$ such that $\epsilon_k <0$ is filled. Then, we measure the energy of the state prepared by doing an inverse FFFT followed by measurement of $Z$ on every site, corresponding to measuring the mode occupation numbers. The total circuit is thus equivalent to a FFFT followed by a reverse FFFT, on top of the initial product state preparation (a barrier is added between the two FFFTs to preclude gate cancellation at compilation). The results of these benchmarks are given in panel b of Fig \ref{fig:fermionic_swaps}, where we plot both the energy measured (bullet points) and the infidelity obtained (bars). These can be compared to the two-qubit gate count and depth for each of these circuits in panel c of Fig \ref{fig:fermionic_swaps}. Overall, we observe that the $\log^2(N)$ compilation \cite{constantinides2025low} gives the lowest energies and infidelities. The graph decimation technique displays close performance. The non-unitary approach that involves measurement and feed-forward performs significantly worse. Finally, it is worth noting that the local CZ implementation, while giving the worst performance among the unitary approaches, performs better than what its relatively large two-qubit gate count and depth could have suggested. This might be  attributed to lower transport time of ions. Following this benchmark we will use the $\log^2(N)$ compilation for the remainder of the paper.

The sensitivity of expectation values to hardware noise is an important aspect of an algorithm. It is for example known that in many settings of Hamiltonian dynamics, errors ``dilute" with system size \cite{schiffer2024quantum,granet2025dilution,chertkov2026robustness,gonzalez2026stability}. In Fig \ref{fig:histogram_fft} we study the effect of hardware noise on the state preparation with FFT, when using the $\log^2(N)$ compilation. In the ideal setting, all the modes between $-\pi/2$ and $\pi/2$ should have probability $1$ of being measured as occupied, and all other modes should be measured as empty. We observe that with hardware noise, the distribution is damped towards $1/2$ (which would be obtained with a completely depolarized state) but strongly remains step-like. The ground state prepared with FFT is thus far from being a thermal state, contrary to what is obtained with other methods \cite{granet2025adiabatic}. Moreover, we also observe that the distribution of the number of incorrect modes measured is far from being a Poisson distribution. This means that each error occurring must violate several modes at the same time, suggesting a rather strong sensitivity to noise of this ground state preparation with FFT. This suggests that error dilution does not hold for the FFT, although a systematic study with increasing system sizes would be required to draw a conclusion.

\section{Implementation of our protocol on quantum hardware} We now present an implementation of our method for measuring spectral functions on Quantinuum's H2-2 ion-trap quantum computer. We consider a $N=27$-site system, together with a $27$-site environment, using $54$ qubits in total. The system + environment is encoded with a Jordan-Wigner transformation, with a fermionic mode ordering that alternates between system and environment, namely $c_1,d_1,c_2,d_2,c_3,d_3,...$. As a demonstration our technique, we consider the one-dimensional Hamiltonian
\begin{equation}\label{Hamiltonian}
    H_{\rm sys}=-\sum_{j=1}^N c_j^\dagger c_{j+1}+c_{j+1}^\dagger c_j\,,
\end{equation}
with periodic boundary conditions. This free-fermion Hamiltonian allows for an exact simulation of the $54$-qubit circuit to benchmark our method. We prepare the ground state of this Hamiltonian using a $27$-qubit FFFT on the system. Then we set a coupling $\epsilon=0.5$, perform $10$ Trotter steps of time evolution of the full Hamiltonian $H$ with time step ${\rm d}t=0.5$. At the end of the time evolution, we perform a $2$-way inverse interleave on the $54$-qubit system+environment in order to bring all the environment fermionic modes next to each other. We then perform a $27$-qubit FFFT on the environment. To compute the entire spectral function $A(k,\omega)$, we run two different settings, one where the environment is initially empty, and one where it is full. The fermions in the environment are added after the ground state preparation, and in order to preserve the fermionic statistics, this requires acting with $Z_j$ on every other system qubit, starting with the first one. Then we measure the environment in the $Z$ basis.

The hardware results are shown in Fig \ref{fig:hardware}. A smoothed version with a bicubic interpolation is shown in Fig \ref{fig:conceptual}. In panel a of Fig \ref{fig:hardware}, we display the expectation value of $n(k)$ as a function of $k$ at $\omega=0$, corresponding to the measured value of $A^\pm(k,0)$ and the full spectral function $A(k,0)$, comparing with the exact noiseless simulation of the circuits. We observe a general attenuation of the signal due to hardware noise (for a completely depolarized state the expectation value on $A^\pm$ would be $1/2$). However, the two peaks are still clearly visible on $A(k,\omega)$ by several standard deviations. The ``noise floor" that appears, roughly independent of momentum $k$, can be interpreted by the locality of noise: errors occur on single or pairs of qubits, which results in a flat effect in momentum space. Next in panel b we show a color plot of the obtained intensity, comparing again to noiseless values that can be computed because of the free fermion structure. We observe very good agreement of the shape of the band. We see again a general attenuation towards signal equal to $1$ due to hardware noise. However, besides the loss of contrast, hardware noise has no effect on the shape of the band, and thus physical quantities like the effective electron mass or the group velocity are effectively unbiased.

\section{Discussion} We have introduced a method to compute spectral functions on quantum computers that require $\mathcal{O}(N)$ fewer shots than previous approaches based on computing dynamical correlations. This translates in a factor $\mathcal{O}(N)$ improvement in runtime. The method consists in a direct modeling of the interaction between the system and an environment, as would occur in actual ARPES experiments (albeit in a simplified way). Although the method requires to double the number of qubits and comes with a two-qubit gate overhead due to environment interactions and need for fermionic Fourier transform, the sparing of a factor $\mathcal{O}(N)$ in the number of shots provides a crucial runtime improvement for ion-trap quantum computers, where shot time is comparatively slow. We demonstrated our algorithm on Quantinuum's H2-2 ion-trap hardware with spectral functions of one-dimensional fermionic models on $27$ sites, requiring $54$ qubits. This required developing a radix-$3$ fermionic Fourier transform that we benchmarked on hardware. The shot reduction brought by our algorithm increases with system size, and will become fundamental for ion-trap quantum computers in the utility scale, where we expect to deal with $10$ to $100$ times larger system sizes.

\section*{Acknolwedgments} We thank Eli Chertkov, Juan Pedersen and Matthew DeCross for comments on the draft. E.G. acknowledges support by the Bavarian Ministry of Economic Affairs, Regional Development and Energy (StMWi) under project Bench-QC (DIK0425/01).


%



\appendix
\section{Free fermion case.}
\subsection{Derivation}
In this section, we treat the exactly solvable case where the system Hamiltonian is given by a free-fermion hopping Hamiltonian
\begin{equation}
    H_{\rm sys}=\nu\sum_{j=1}^N c_j^\dagger c_{j+1}+c_{j+1}^\dagger c_j\,.
\end{equation}
We introduce
\begin{equation}
    \gamma_j=\left(\begin{matrix}
        c_j\\ d_j
    \end{matrix} \right)\,,\qquad \gamma_j^\dagger=\left(\begin{matrix}
        c_j^\dagger& d_j^\dagger
    \end{matrix} \right)\,,
\end{equation}
which satisfy canonical anticommutation relations $\{\gamma_j,\gamma_\ell^\dagger\}=\delta_{j,\ell}$. The full Hamiltonian $H= H_{\rm sys}+\epsilon H_{\rm int}+\omega H_{\rm env}$ can be written as
\begin{equation}
\begin{aligned}
    H=&\sum_{j=1}^N \nu \gamma_{j+1}^\dagger \left(\begin{matrix}
        1&0\\ 0&0
    \end{matrix} \right)\gamma_j+\nu\gamma_{j}^\dagger \left(\begin{matrix}
        1&0\\ 0&0
    \end{matrix} \right)\gamma_{j+1}\\
    &+\frac{\epsilon}{2} \gamma_{j}^\dagger \left(\begin{matrix}
        0&1\\ 1&0
    \end{matrix} \right)\gamma_j+\omega\gamma_j^\dagger \left(\begin{matrix}
        0&0\\ 0&1
    \end{matrix} \right)\gamma_j\,.
\end{aligned}
\end{equation}
Introducing then the momentum space fermions
\begin{equation}
    \gamma(k)=\frac{1}{\sqrt{N}}\sum_{j=1}^N e^{-ijk}\gamma_j
\end{equation}
for $k\in K=\{2\pi n/N, n=0,...,N-1\}$
we have
\begin{equation}
    H=\sum_{k\in K} \gamma^\dagger(k)M(k) \gamma(k)\,,
\end{equation}
with the matrix
\begin{equation}
    M(k)=\left(\begin{matrix}
        2\nu\cos k&\epsilon/2 \\ \epsilon/2 &\omega
    \end{matrix} \right)\,.
\end{equation}
Its eigenvalues are $e_k\mp \eta_k$ with
\begin{equation}
   \eta_k=\frac{1}{2}\sqrt{\left(\omega-2\nu\cos k\right)^2+\epsilon^2 }\,,\qquad e_k=\frac{\omega+2\nu\cos k}{2}\,,
\end{equation}
and corresponding eigenvectors are 
\begin{equation}
   \left(\begin{matrix}
        \cos \theta_k\\ \sin\theta_k
    \end{matrix}\right)\,,\qquad \left(\begin{matrix}
        \sin \theta_k\\ -\cos\theta_k
    \end{matrix}\right)\,,
\end{equation}
where
\begin{equation}
    \tan \theta_k=\frac{1}{\epsilon }\left(\omega-2\nu\cos k-\sqrt{\left(\omega-2\nu\cos k\right)^2+\epsilon^2 }\right)\,.
\end{equation}
We now define the fermions
\begin{equation}
    \begin{aligned}
        \alpha_k^\dagger&= \gamma^\dagger(k)\cdot  \left(\begin{matrix}
        \cos \theta_k\\ \sin\theta_k
    \end{matrix}\right)=\cos \theta_k c^\dagger(k)+\sin \theta_k d^\dagger(k)\\
    \beta_k^\dagger&= \gamma^\dagger(k)\cdot  \left(\begin{matrix}
        \sin \theta_k\\ -\cos\theta_k
    \end{matrix}\right)=\sin \theta_k c^\dagger(k)-\cos \theta_k d^\dagger(k)\,,
    \end{aligned}
\end{equation}
which also satisfy canonical anticommutation relations $\{\alpha_k,\alpha_p^\dagger\}=\delta_{k,p}$, $\{\beta_k,\beta_p^\dagger\}=\delta_{k,p}$, $\{\alpha_k,\beta_p^\dagger\}=0$, all other anticommutation relations being $0$. In terms of them, we have
\begin{equation}
    H=\sum_{k\in K} (e_k-\eta_k)\alpha^\dagger_k \alpha_k+(e_k+\eta_k)\beta_k^\dagger \beta_k\,.
\end{equation}
Hence, the eigenvectors of $H$ are built by applying some $\alpha^\dagger_k, \beta_k^\dagger$ on the vacuum state.

Let us now start from an eigenstate of $H_{\rm sys}$, obtained for $\epsilon=0$, and evolve the system with $H$ for $\epsilon> 0$. We would like to measure the mode occupation number in momentum space in the auxiliary space. This is
\begin{equation}
\begin{aligned}
    n_k=&d^\dagger(k)d(k)\\
    &=\sin^2 \theta_k \alpha^\dagger_k \alpha_k + \cos^2\theta_k \beta_k^\dagger \beta_k\\
    &-\cos \theta_k \sin \theta_k (\beta_k^\dagger \alpha_k + \alpha_k^\dagger \beta_k)\,.
\end{aligned}
\end{equation}
After time $t$, it is
\begin{equation}
\begin{aligned}
    n_k(t)=&\sin^2 \theta_k \alpha^\dagger_k \alpha_k + \cos^2\theta_k \beta_k^\dagger \beta_k\\
    &-\cos \theta_k \sin \theta_k (\beta_k^\dagger \alpha_k e^{2i\eta_k t} + \alpha_k^\dagger \beta_k e^{-2i\eta_k t})\,.
\end{aligned}
\end{equation}
In the initial state, we have
\begin{equation}
\begin{aligned}
 &\langle \alpha_k^\dagger \alpha_k\rangle=\cos^2\theta_k \rho_k\\
 &\langle \beta_k^\dagger \beta_k\rangle=\sin^2\theta_k \rho_k\\
 &\langle \beta_k^\dagger \alpha_k\rangle=\langle \alpha_k^\dagger \beta_k\rangle=\sin\theta_k\cos\theta_k \rho_k\,,
\end{aligned}
\end{equation}
where $\rho_k$ is the mode occupation number in the initial state. Hence we find
\begin{equation}
    \langle n_k(t)\rangle=\sin^2(2\theta_k) \sin^2(\eta_k t)\rho_k\,.
\end{equation}
Explicitly, this is
\begin{equation}\label{formula1d}
\begin{aligned}
    \langle n_k(t)\rangle&=\frac{\epsilon^2\sin^2(t\Omega)}{\epsilon^2+\left(\omega-2\nu\cos k\right)^2}\rho_k \\
    \Omega&=\frac{1}{2}\sqrt{\epsilon^2+\left(\omega-2\nu\cos k\right)^2}\,.
\end{aligned}
\end{equation}
\subsection{Limiting cases}
At small $\epsilon$ and fixed $\nu$, this is
\begin{equation}
    \langle n_k(t)\rangle=\frac{\epsilon^2}{(\omega -2\nu\cos k)^2} \sin^2\left( \frac{\omega -2\nu\cos k}{2}t\right)\rho_k  +\mathcal{O}(\epsilon^4)\,.
\end{equation}
Let's compare to the generic formula \eqref{result} in the main text obtained at small $\epsilon$. We have the Green function
\begin{equation}
    G(k,\omega)=\sum_{j=1}^N \int_{-\infty}^\infty \D{t} \langle c_j^\dagger(t) c_0 \rangle e^{ikj}e^{-i\omega t}\,.
\end{equation}
The expectation value within the initial state considered is
\begin{equation}
    \langle c_j^\dagger(t) c_0 \rangle=\frac{1}{N}\sum_{k\in K} e^{-ikj} e^{2it\nu \cos k} \rho_k\,.
\end{equation}
Hence
\begin{equation}
    G(k,\omega)= 2\pi \rho_k \delta(2\nu\cos k-\omega)\,.
\end{equation}
So formula \eqref{result} in the main text agrees with \eqref{formula1d} indeed.

\section{Strong coupling expansion.}
\subsection{Leading order}
We now consider the Hamiltonian
\begin{equation}
    H= \nu H_{\rm sys}+\epsilon H_{\rm int}+\omega H_{\rm env}
\end{equation}
for generic $H_{\rm sys}$. We perform a strong coupling expansion, namely an expansion in $\nu$, at fixed $\epsilon$. At order $\nu^0$, we have
\begin{equation}
    U(t)=e^{it(\epsilon H_{\rm int}+\omega H_{\rm env})}+\mathcal{O}(\nu)\,,
\end{equation}
and so we get at this order
\begin{equation}
\begin{aligned}
&\partial_t c_{j}=i\frac{\epsilon}{2} d_j\\
    &\partial_t d_j=i\left(\frac{\epsilon}{2} c_j+\omega d_j\right)\,.
\end{aligned}
\end{equation}
Hence, up to an irrelevant $e^{i\omega t/2}$ global phase
\begin{equation}
    \left(\begin{matrix}
        c_j(t)\\d_j(t)
    \end{matrix}\right)=\exp \left(\frac{it}{2}\left(\begin{matrix}
        -\omega & \epsilon \\ \epsilon & \omega
    \end{matrix}\right)\right)\left(\begin{matrix}
        c_j\\d_j
    \end{matrix}\right)\,.
\end{equation}
The exponential is
\begin{equation}
\begin{aligned}
    &\exp \left(\frac{it}{2}\left(\begin{matrix}
        -\omega & \epsilon \\ \epsilon & \omega
    \end{matrix}\right)\right) \\
    &=\cos(t\Omega_0)\left(\begin{matrix}
        1 & 0 \\ 0 & 1
    \end{matrix}\right)+i\frac{\sin(t\Omega_0)}{\sqrt{\omega^2+\epsilon^2}}\left(\begin{matrix}
        -\omega & \epsilon \\ \epsilon & \omega
    \end{matrix}\right)\,,
\end{aligned}
\end{equation}
with
\begin{equation}
    \Omega_0=\frac{1}{2}\sqrt{\omega^2+\epsilon^2}\,.
\end{equation}
Hence we find
\begin{equation}\label{coeff}
    \begin{aligned}
        &c_j(t)=a(t) c_j +a'(t) d_j\\
        &d_j(t)=a(t)^* d_j+a'(t) c_j\,,
    \end{aligned}
\end{equation}
with
\begin{equation}
\begin{aligned}
    &a=\cos(t\Omega_0)-i\sin(t\Omega_0)\frac{\omega}{\sqrt{\omega^2+\epsilon^2}}\\
    &a'=i\sin(t\Omega_0)\frac{\epsilon}{\sqrt{\omega^2+\epsilon^2}}\,.
\end{aligned}
\end{equation}
We obtain thus
\begin{equation}\label{strong0}
    \langle n_k(t)\rangle=\sin^2(t\Omega_0)\frac{\epsilon^2}{\omega^2+\epsilon^2} \langle E| c^\dagger(k) c(k) |E\rangle+\mathcal{O}(\nu)\,.
\end{equation}

\section{Optimized interleave operations using graph decimation.}
In this section we describe how to apply the graph decimation approach to implement the CZ circuits appearing in fermionic permutation networks. We also present explicit applications of the approach to the 3-way interleave on 9 and 27 qubits.

Any CZ circuit can be defined by a graph $G$ consisting of nodes (qubits) $i\in V$ and edges $(i,j) \in E\subset [V]^2$, which are identified by unordered pairs of nodes, such that,
\begin{equation}
    U_G = \prod_{(i,j) \in E} {\rm CZ}_{i,j}.
\end{equation}
The graph decimation approach was originally designed to prepare the graph state $U_G\ket{+\dots +}$ \cite{doherty2026fast}, but it can be straightforwardly modified to implement $U_G$ on arbitrary input states. The goal is to find an optimal sequence of gates that maps $U_G$ to the identity matrix, where each gate acts via right and/or left multiplication. Reversing this sequence then produces an optimized circuit realization of $U_G$. We consider operations that map $U_G$ to $U_{G'}$ where $G'$ is a new graph, ideally with fewer edges than $G$. Thus, the circuit synthesis problem can be phrased as finding a minimal sequence of graph operations that completely disentangle the initial graph $G$, hence the name graph decimation.

We consider the following three types of operations at each step of graph decimation,
\begin{align}
     U_G {\rm CZ}_{i,j} &=  U_{G \oplus \{(i,j)\}} \label{eq:rule1} \\
     {\rm CX}_{i,j} U_G {\rm CX}_{i,j} &= U_{G \oplus \{(i,k): k\in n(j)\}} \label{eq:rule2} \\
     {\rm CX}_{i,j} U_G {\rm CY}_{i,j} &= U_{G \oplus \{(i,k): k\in n(j) \cup \{j\}\}} \label{eq:rule3},
\end{align}
where ${\rm CY} = (\sqrt{Z}\otimes \sqrt{X}){\rm CZ}(I\otimes \sqrt{X}^\dagger)$ is the controlled-$i{\rm Y}$ gate, $n(i) = \{j \,\vert\, (i,j)\in E\}$ is the neighborhood of node $i$, and $G \oplus \Gamma $ for a set of edges $\Gamma$ is a graph whose edges are obtained by taking the symmetric difference of the edges in $G$ with those in $\Gamma$. The first relation follows from the definition of $U_G$. The second relation follows from a simple gate identity describing how ${\rm CZ}$ gates transform under conjugation by ${\rm CX}$ gates. The third relation is simply the composition of the first two, since ${\rm CY} = {\rm CZ} \circ {\rm CX}$. The cost of the three operations, defined as the number of two-qubits gates and denoted as $C$, is 1, 2, and 2, respectively.

Let $|G|$ denote the number of edges in $G$. Note that $|G|$ represents an upper bound on the number of two-qubit gates needed to implement $U_G$, since we can simply remove the edges of $G$ one-by-one using Eq.~\eqref{eq:rule1}. However, by judiciously applying the other rules, we can often implement $U_G$ using fewer gates. To find such minimal gate sequences, we proceed in the following greedy fashion. At each step, we consider all possible operations from Eqs.~\eqref{eq:rule1}-\eqref{eq:rule3} on all pairs of nodes and choose that one that minimizes $|G'| - |G| + C$ where $G'$ is the new graph obtained after applying that operation. (That is, we choose the operation that reduces the edge count of $G$ the most, where the operations in Eqs.~\eqref{eq:rule2} and \eqref{eq:rule3} get a penalty of 1 since they cost one more two-qubit gate than Eq.~\eqref{eq:rule1}.) We can simultaneously optimize for circuit depth by applying a tunable penalty to each operation that would start a new circuit layer. 

The procedure ends once $G$ has no edges remaining, meaning $U_G = I$, and the final circuit for $U_G$ is obtained by applying the obtained sequence of operations in reverse, starting from $I$. Note that it is always possible to remove at least one edge by applying a ${\rm CZ}$ gate, so the procedure will always terminate. 

Above, we described a simple greedy approach to the graph decimation problem. However, as shown in Ref.~\cite{doherty2026fast}, we can obtain even better gate sequences by applying more sophisticated methods such as beam search or reinforcement learning. The sequences described below were obtained using the simple greedy method, which was found to perform comparably well for system sizes $< 30$ ~\cite{doherty2026fast}.

To conclude, we present the results of graph decimation when applied to CZ circuits corresponding to the $3$-way inverse interleave operation on 9 and 27 qubits. As in Fig \ref{fig:fermionic_swaps}, the permutation of the qubits corresponding to the inverse interleave permutation has to be applied before these circuits.

\onecolumngrid

\subsection{$3$-way inverse interleave on $9$ qubits}
\begin{verbatim*}
CX(6,3)CZ(7,2)

CZ(6,4)CZ(3,2)CZ(7,5)

CZ(1,3)CZ(6,5)CZ(2,4)

CX(6,3)
\end{verbatim*}

\subsection{$3$-way inverse interleave on $27$ qubits}
\begin{verbatim*}
CX(18,19)CX(8,7)CX(20,21)CX(6,5)CX(16,17)CX(9,10)CX(15,14)CX(11,12)CX(22,24)CX(3,2)

CX(19,21)CX(5,4)CX(7,17)

CZ(20,12)CZ(10,2)CZ(15,6)CZ(25,16)CZ(18,9)CZ(11,3)CZ(22,14)

CZ(13,7)CZ(23,17)CZ(21,4)CZ(19,12)CZ(22,5)CZ(14,6)CZ(18,10)CZ(25,8)CZ(9,1)CZ(20,3)

CZ(21,17)CZ(14,7)CZ(10,4)CZ(13,5)CZ(23,15)CZ(19,2)CZ(18,1)CZ(16,8)CZ(20,11)

CZ(12,7)CZ(21,13)CZ(24,17)CZ(23,6)

CX(3,2)

CZ(21,14)CZ(12,4)

CX(7,17)CX(22,24)

CZ(25,17)CZ(10,7)

CX(5,4)CX(19,21)CX(11,12)CX(15,14)

CX(9,10)CX(16,17)CX(6,5)CX(20,21)CX(8,7)CX(18,19)
\end{verbatim*}

\end{document}